# Conocimientos sobre astronomía en estudiantes de educación secundaria en Colombia: una evaluación desde la fundación AstrodidaXis

## Astronomy knowledge in secondary school students in Colombia: an evaluation from the AstrodidaXis

Daniel Alejandro Valderrama[1]
Juan Camilo Guzmán Rodríguez[2]
Julián David Umbarila Benavides[3]
*UPTC*

Néstor Eduardo Camino[4]
*CONICET*

Lorena María González Pardo[5]
*ITINAR*

[1] daniel.valderrama@uptc.edu.co
https://orcid.org/0000-0002-3360-3890
[2] juan.guzman05@uptc.edu.co
https://orcid.org/0000-0002-5993-3775
[3] julian.umbarila@uptc.edu.co
https://orcid.org/0000-0002-0601-1427
[4] nestor.camino@fhcs.unp.edu.ar
https://orcid.org/0000-0003-1091-5741
[5] lmgonzalezp@itinar.edu.co
https://orcid.org/0009-0005-3264-1712

243

**Resumen**

Se realizó un efectuó exploratorio del conocimiento astronómico de 241 estudiantes de educación secundaria en Boyacá, Colombia, miembros de la red AstrodidaXis, empleando una metodología cualitativa, hermenéutica y exploratoria. Este análisis se desarrolló a través de un cuestionario alineado con los estándares de aprendizaje del Ministerio de Educación Nacional, como base para identificar áreas de fortaleza y debilidad en conceptos de astronomía. Se destacó la necesidad de mejorar la comprensión de diversos temas, incluyendo las fuerzas fundamentales del universo y el origen de los elementos químicos desde la didáctica de la astronomía. Dichos hallazgos contribuyeron a la construcción de proyecciones para la investigación y el desarrollo de la enseñanza de la astronomía en Boyacá, impulsando el progreso científico, tecnológico y social de la región.

**Palabras clave:** Astronomía, educación, enseñanza, divulgación científica, habilidades

**Abstract**

An exploratory analysis of the astronomical knowledge of 241 secondary education students in Boyacá, Colombia, members of the AstrodidaXis network, was carried out using a qualitative, hermeneutic and exploratory methodology. This analysis was developed through a questionnaire aligned with the learning standards of the Ministry of National Education, as a basis for identifying areas of strength and weakness in astronomy concepts. The need to improve the understanding of various topics was highlighted, including the fundamental forces of the universe and the origin of chemical elements from the teaching of astronomy. These findings contributed to the construction of projections for the research and development of astronomy teaching in Boyacá, promoting the scientific, technological and social progress of the region.

**Keywords:** Astronomy, education, teaching, scientific dissemination, skills

**Introducción**

La Astronomía, como ciencia, ha evolucionado a partir de observaciones que han dado lugar a conceptos fundamentales en ciencias exactas y naturales, como la matemática, la física y la química. Estas conexiones, junto con los avances tecnológicos y sus impactos sociales a lo largo de la historia, se han registrado en diferentes períodos de la civilización (Cardona, 2020; Cecil-Dolmange, 2022). A partir de esta base, se ha reconocido el potencial de la Astronomía para integrar conocimientos y contextualizarlos en fenómenos cotidianos observables, fomentando el uso de habilidades científicas que influyen en la toma de decisiones y pueden contribuir al desarrollo social.

Estos logros subrayan la relevancia de la investigación en la enseñanza específica de la Astronomía como una disciplina emergente, que se beneficia del enfoque didáctico de las ciencias naturales (Camino, 2021). Este proceso ha experimentado avances significativos en Latinoamérica, especialmente en la comprensión de conceptos astronómicos, la adopción de enfoques interdisciplinarios y su integración en los planes de estudio. En Uruguay, se enseña bajo el título "Ciencias de la Tierra y el Espacio" (Ganón & Fernández, 2008), mientras que, en Brasil, se incluye en la unidad de Astronomía dentro de Ciencias Naturales (Santos et al., 2022). Por otro lado, en Argentina, la presencia de la Astronomía en el currículo es más limitada, con solo 10 jurisdicciones que la incluyen, principalmente en el ciclo básico y en la formación docente de profesores en Física (Camino et al., 2021; Camino et al., 2022).

Es relevante señalar que dicha incorporación, visibilizada anteriormente, no siempre garantiza un mejor desarrollo de su enseñanza, ya que enfrenta desafíos como la comprensión del conocimiento astronómico, la formación limitada de maestros y la persistencia de concepciones alternativas en estudiantes y docentes. En Colombia y otros países latinoamericanos, ha habido un aumento en la investigación posgradual en la enseñanza de la Astronomía en las últimas décadas, aunque su impacto se concentra en áreas geográficas específicas y la enseñanza de la disciplina sigue siendo limitada en muchas instituciones (Valderrama et al., 2021a; Vargas et al., 2021).

245

En cuanto a las políticas curriculares, en Colombia estas se apoyan en dos documentos clave: los Derechos Básicos de Aprendizaje y los Estándares Básicos de Aprendizaje. Estos documentos integran temas astronómicos en asignaturas como Ciencias Naturales y Ciencias Sociales (como se muestra en la tabla 2 de la metodología). Sin embargo, no se ha evaluado el progreso en el aprendizaje de Astronomía, especialmente en términos de comprensión y asimilación de conceptos respaldados por la comunidad científica.

En el contexto específico de Boyacá, se han observado esfuerzos innovadores en la enseñanza de la Astronomía en la educación primaria. Además, se llevó a cabo el primer Workshop sobre Enseñanza de la Astronomía en 2021, el cual dio lugar a la creación de la red AstrodidaXis, una unión entre la universidad, docentes de primaria y secundaria, y estudiantes en formación con el objetivo de promover la divulgación y formación en Astronomía en la región.

Es así como surge la necesidad de evaluar los conocimientos astronómicos de los estudiantes en Boyacá, con el fin de fortalecer la investigación e innovación didáctica en Astronomía basada en la experiencia regional. La pregunta de investigación que se plantea es: ¿Qué conceptos científicos sobre Astronomía poseen los estudiantes de la fundación red AstrodidaXis en el departamento de Boyacá?

**Metodología**

Esta investigación emplea un enfoque cualitativo y un método de investigación hermenéutico para identificar construcciones conceptuales en sujetos sociales (Hernández-Sampieri & Mendoza, 2018). El alcance es exploratorio, ya que busca generar ideas para futuras investigaciones en la didáctica de la Astronomía, sin pretender describir de manera exhaustiva la realidad bajo estudio.

Participantes

Este estudio se desarrolló en el marco de las actividades de la red AstrodidaXis, integrando a 241 jóvenes con edades entre 10 y 19 años, de los cuales 120 eran de sexo femenino y 121 de sexo masculino. Los participantes eran oriundos de diferentes municipios del departamento de Boyacá, Colombia, y cursaban distintos grados de educación secundaria, como se observa en la Tabla 1.

Tabla 1. Grado de escolaridad y edades de los participantes

| Grado de escolaridad | Rango de edades (años) | Cantidad de estudiantes |
| --- | --- | --- |
| 6° | 13-14 | 3 |
| 7° | 13-15 | 25 |
| 8° | 10-19 | 50 |
| 9° | 13-18 | 39 |
| 10° | 13-18 | 64 |
| 11° | 14-18 | 60 |

Fuente: Elaboración propia.

Instrumento

Se diseñó un cuestionario de 11 preguntas de opción múltiple sobre Astronomía, en el que se abordaron cuatro temas principales basados en los estándares y derechos de aprendizaje sugeridos por el Ministerio de Educación Nacional (2011a) y el Ministerio de Educación de Colombia (2016). El cuestionario fue validado por tres expertos: dos de ellos investigadores locales en Astronomía y uno internacional. Además, fue diseñado por los autores del estudio.

Tabla 2. Tópicos principales para las preguntas según los Estándares básicos por competencias para Ciencias Naturales y Ciencias Sociales (Ministerio de Educación Nacional, 2011a) y los Derechos Básicos de aprendizaje (Min educación, 2016)

| Categoría | Estándares Básicos por competencias | | Derechos Básicos de aprendizaje (DBA C.N) y (DBA C.S) | | | | Preguntas |
|---|---|---|---|---|---|---|---|
| | **Estándar** | **Grado** | **Derecho** | **Evidencia** | **Grado** | **CN/CS** | |
| Relaciones Tierra, Luna, Sol | Registro el movimiento del Sol, la Luna y las estrellas en el cielo, en un período de tiempo | 1° a 3° | Comprende que el fenómeno del día y la noche se debe a que la Tierra rota sobre su eje y en consecuencia el sol solo ilumina la mitad de su superficie | Registra y realiza dibujos de las sombras que proyecta un objeto que recibe la luz del Sol en diferentes momentos del día, relacionándolas con el movimiento aparente del Sol en el cielo. | 4° | Ciencias Naturales | P1, P2, P3 |
| | Reconozco diversas formas de representación de la Tierra | | | Explica cómo se producen el día y la noche por medio de una maqueta o modelo de la Tierra y del | | | |

| | | | | | | |
|---|---|---|---|---|---|---|
| | Identifico formas de medir el tiempo (horas, días, años...) y las relaciono con las actividades de las personas. | | | Sol. Comprende que el fenómeno del día y la noche se debe a que la Tierra rota sobre su eje y en consecuencia el sol sólo ilumina la mitad de su superficie | | |
| | Describo los principales elementos del sistema solar y establezco relaciones de tamaño, movimiento y posición. | 4° y 5° | Comprende que las fases de la Luna se deben a la posición relativa del Sol, la Luna y la Tierra a lo largo del mes. | Realiza observaciones de la forma de la Luna y las registra mediante dibujos, explicando cómo varían a lo largo del mes | | |
| | Relaciono el movimiento de traslación | | | Predice cuál sería la fase de la Luna que un observador | | |
| | con los cambios climáticos | | | vería desde la Tierra, dada una cierta posición relativa entre la Tierra, el Sol y la Luna | | | |
| Sistema Solar | Explico el modelo planetario desde las fuerzas gravitacionales. Relaciono masa, peso y densidad con la aceleración de la gravedad en distintos puntos del sistema solar | 6° y 7° | Comprende que existen diversas explicaciones y teorías sobre el origen del universo en nuestra búsqueda por entender que hacemos parte de un mundo más amplio | Explica los elementos que componen nuestro sistema solar: planetas, estrellas, asteroides, cometas y su relación con la vida en la Tierra. | 6° | Ciencias Sociales | P4 |
| Estrellas | Describo el proceso de formación y | 6° y 7° | | | | | P5, P6, P7, P8 y P9 |

| | | | | | | |
|---|---|---|---|---|---|---|
| | extinción de estrellas. | | | | | |
| Cosmología | Establezco relaciones entre el modelo del campo gravitacional y la ley de gravitación universal. | 10° y 11° | Comprende que existen diversas explicaciones y teorías sobre el origen del universo en nuestra búsqueda por entender que hacemos parte de un mundo más amplio | Interpreta diferentes teorías científicas sobre el origen del universo (Big Bang, inflacionaria, multiuniversos), que le permiten reconocer cómo surgimos, cuándo y por qué | 6° | Ciencias sociales | P10, P11 |

Fuente: Adaptado de Valderrama et al., (2021b)

**Consideraciones éticas**

El cuestionario se llevó a cabo de noviembre de 2021 a mayo de 2022, con la participación voluntaria e informada de docentes y estudiantes de la fundación Red AstrodidaXis. Se utilizó la plataforma Google Forms y se garantizó el anonimato de los participantes al no recopilar datos identificativos. Se siguieron las normas de protección de la información y, en el caso de los estudiantes menores de edad, se informó a los padres de familia sobre los propósitos y la metodología de la investigación.

**Resultados y discusión de resultados**

A continuación, se presentan los resultados de las respuestas obtenidas por cada categoría del cuestionario suministrado.

Relaciones Tierra, Luna, Sol

La interacción gravitacional entre la Tierra y el Sol provoca varios movimientos en nuestro planeta. Los más conocidos son la traslación, el recorrido alrededor del Sol que dura aproximadamente 365,25 días, y la rotación, el giro de la Tierra sobre su eje con un período de 24 horas. También existen otros movimientos menos explorados en la educación, como la precesión, que es un movimiento rotacional de largo período, casi 26.000 años, que realiza el eje de rotación terrestre respecto a la perpendicular al plano orbital, y la nutación, un "bamboleo" del eje de rotación terrestre respecto a su posición media, con un período cercano a los 19 años (Morison, 2008). Estos últimos movimientos son causados por la interacción gravitatoria de la Tierra con la Luna y otros cuerpos del Sistema Solar. Estos movimientos son casi imperceptibles en la observación cotidiana y se plantean en la pregunta 1.

*Pregunta 1*

Para identificar los conocimientos de los estudiantes sobre los movimientos de la Tierra, se utilizó la siguiente pregunta basada en la Figura 1: "es necesario mencionar que los números ubicados en esta corresponden a los movimientos de la Tierra. Seleccione la respuesta en la que los números coincidan con el nombre del movimiento."

a) <u>1 precesión, 2 rotación, 3 nutación, 4 traslación</u>
b) 1 nutación, 2 rotación, 3 precesión, 4 traslación
c) 1 precesión, 2 traslación, 3 nutación, 4 rotación

Figura 1. Movimientos de la Tierra

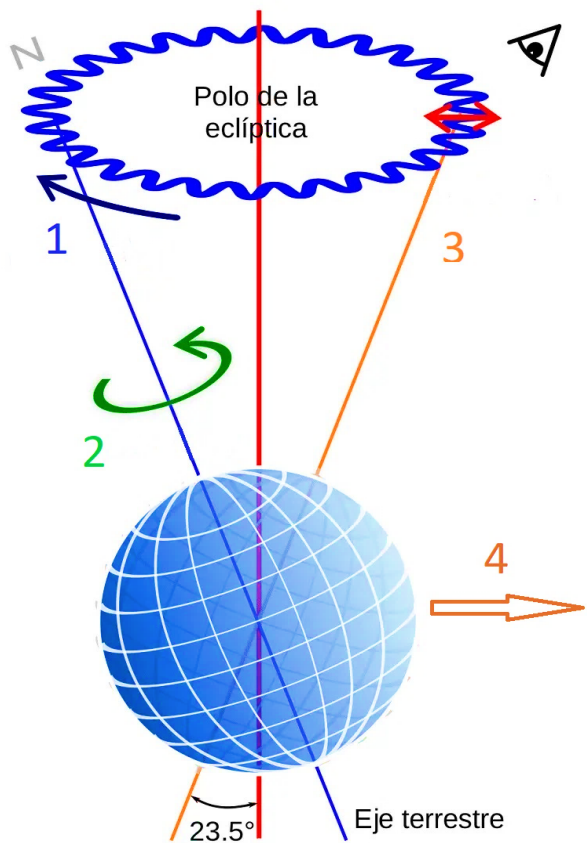

Fuente: Pederzoli, 2011.

Los resultados obtenidos se encuentran resumidos en la Figura 2, donde se evidencia que el 46% de los estudiantes reconoce correctamente los movimientos de la Tierra presentados en

253

la Figura 1. Además, el 26% de los estudiantes reconoce la rotación y traslación, pero presenta confusión respecto a los movimientos de precesión y nutación. Por otro lado, el 28% de los estudiantes no reconoce los posibles movimientos terrestres involucrados en la imagen.

Figura 2. Porcentajes de respuestas de los estudiantes a la pregunta 1 sobre los movimientos de la Tierra

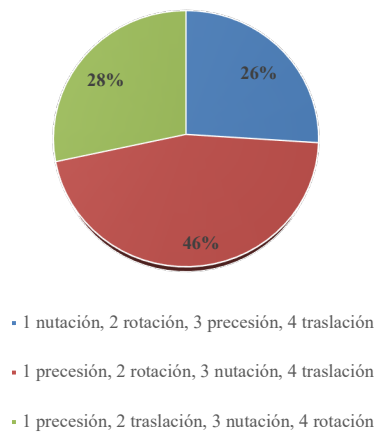

- 1 nutación, 2 rotación, 3 precesión, 4 traslación
- 1 precesión, 2 rotación, 3 nutación, 4 traslación
- 1 precesión, 2 traslación, 3 nutación, 4 rotación

Fuente: elaboración propia.

Si se analiza desde una visión curricular, en Colombia estos contenidos se abordan en educación primaria en los grados 4° y 5°. Sin embargo, se observan dificultades conceptuales en aproximadamente el 58% de los estudiantes. Estos problemas han sido referenciados en estudios similares, los cuales señalan que, a pesar de que los conceptos se abordan en los contenidos de Ciencias Naturales, todavía subsisten dificultades en su apropiación (Delgado-Serrano & Cubilla, 2012; Solbes & Palomar, 2013).

*Pregunta 2*

Los movimientos de la Tierra, como la traslación y la inclinación de su eje (23°27') dan lugar a las estaciones en los diferentes hemisferios del planeta. Esta relación entre los movimientos terrestres y las estaciones del año dan lugar a la pregunta 2.

El movimiento de traslación de la Tierra causa los períodos conocidos como_________que para el caso del hemisferio sur de la Tierra corresponde con los meses de__________.

a) Año, diciembre solsticio de verano, enero, Equinoccio de otoño, febrero, Solsticio de Invierno, marzo Equinoccio de primavera.
b) Estaciones, enero solsticio de verano, febrero Equinoccio de otoño, marzo Solsticio de Invierno, Abril Equinoccio de primavera.
c) Año, diciembre solsticio de verano, marzo Equinoccio de otoño, junio Solsticio de Invierno, septiembre Equinoccio de primavera.
d) <u>Estaciones, diciembre solsticio de verano, marzo Equinoccio de otoño, junio Solsticio de Invierno, Septiembre Equinoccio de primavera.</u>

En la Figura 3, se observa que el 26% de los estudiantes reconocen el cambio estacional con relación al movimiento de traslación de la Tierra, asociando correctamente los conceptos de solsticios y equinoccios con los meses del año en el que se presentan. El 38% de los estudiantes identifica el fenómeno de las estaciones, pero no presenta claridad sobre su ocurrencia en ambos hemisferios y el 36% de los estudiantes no relaciona los conceptos de solsticios y equinoccios correctamente.

Figura 3. Porcentajes de respuestas a la pregunta 2 relacionada con las estaciones del año en el hemisferio sur

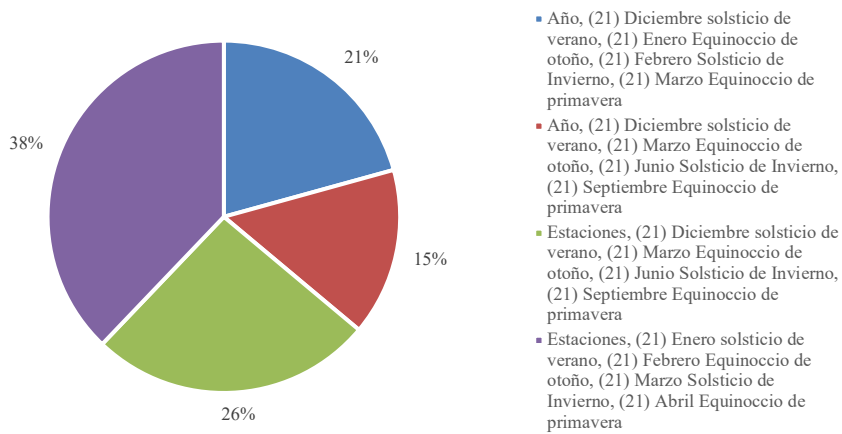

Fuente: Elaboración propia

Se observa que la zona de estudio, ubicada cerca del Ecuador, experimenta estaciones con climas diferentes a las regiones en los trópicos, lo que dificulta la comprensión de conceptos contextualizados en otras latitudes. Esto se ha reflejado en estudios similares en otras partes del país, donde la Arqueoastronomía se está volviendo importante para entender el interés ancestral en solsticios y equinoccios (Buitrago-Sierra, 2019; Romero et al., 2019).

*Pregunta 3*

La Luna, el satélite natural de la Tierra, es uno de los cuerpos celestes más estudiados. Sus características principales incluyen las fases lunares, que son cambios en su iluminación causados por los movimientos relativos en el sistema Tierra-Luna-Sol, con una periodicidad de aproximadamente 29.5 días. Este ciclo comienza con la Luna nueva, donde no refleja luz solar hacia la Tierra; la Luna va aumentando su iluminación hasta llegar al cuarto creciente, luego a la Luna llena, y finalmente disminuye su luminosidad hasta llegar al cuarto menguante, regresando así a la fase de Luna nueva (Seeds & Backman, 2011). A partir de lo anterior se formula la pregunta 3.

En la siguiente imagen (Figura 4) se muestran las fases de la Luna, relacione los nombres con los números respectivos en la imagen.
a) I Luna menguante, II Luna Azul, III Luna creciente, IV Luna gibosa
b) I Cuarto creciente, II Luna llena, III Cuarto menguante, IV Luna Nueva
c) I Luna menguante, II Luna Llena, III Luna Nueva, IV Luna creciente.
d) I Cuarto de menguante, II Luna nueva, III cuarto de creciente y IV Luna llena

256

Figura 4. Imagen fases de la Luna

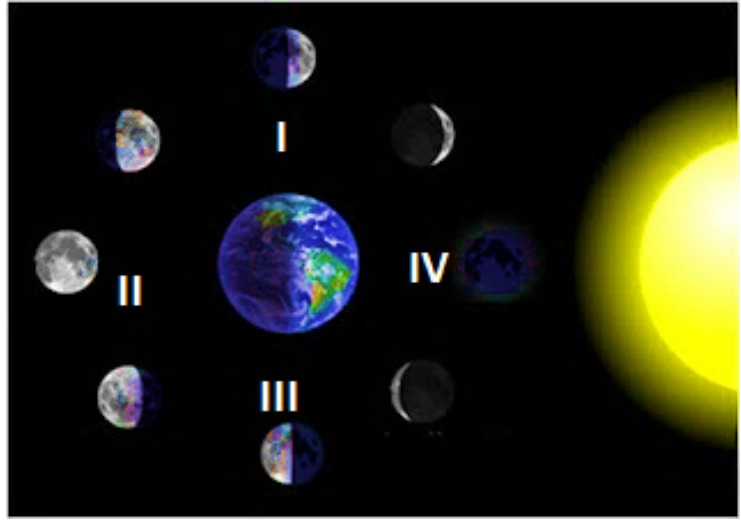

Fuente: Colegio Molière (2013).

En este sentido, la Figura 5 muestra que el 67% de los estudiantes reconocen correctamente las fases lunares. El 10% confunde la Luna llena con la Luna nueva, el 18% confunde la Luna nueva con el cuarto creciente, mientras que el 5% no reconoce ninguna de las fases.

Figura 5. Sistematización de las respuestas para la pregunta 3

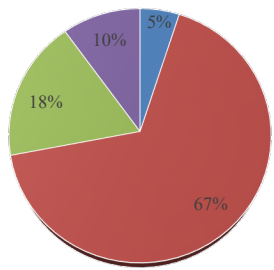

- I Luna menguante, II Luna Azul, III Luna creciente, IV Luna gibosa
- I Cuarto creciente, II Luna llena, III Cuarto menguante, IV Luna Nueva
- I Luna menguante, II Luna Llena, III Luna Nueva, IV Luna creciente.
- I Cuarto de menguante, II Luna nueva, III cuarto de creciente y IV Luna llena

Fuente: Elaboración propia.

257

Es necesario destacar que el estudio de las fases de la Luna aparece en los estándares educativos de Colombia para los primeros grados. Esto se refleja en que la mayoría de los estudiantes reconocen estas fases. Aunque no hay propuestas específicas para enseñar este tema en Boyacá, se observan procesos innovadores en otras regiones del país, que incluyen enfoques socioculturales y actividades lúdicas (Díaz-Moncada, 2019; Cerón-Quiceno et al., 2019).

El sistema solar

El sistema planetario en el que se encuentra la Tierra es denominado sistema solar y reúne todos los objetos que orbitan alrededor del Sol. Se compone de ocho planetas, clasificados en terrestres: Mercurio, Venus, Tierra y Marte, y gigantes gaseosos: Júpiter, Saturno, Urano y Neptuno (Trigo, 2001). Lo anterior dio lugar a la formulación de la pregunta 4.

*Pregunta 4*

En el sistema solar se pueden identificar los siguientes cuerpos celestes: Mercurio, Venus, Tierra, Marte, Júpiter, Saturno, Urano, Neptuno, los cuales se pueden clasificar de acuerdo con su estado de la materia dominante en __________ y _____________".
a) Planetas y asteroides
b) Planetas terrestres y líquidos
c) <u>Planetas terrestres y gaseosos</u>

Conforme se aprecia en la Figura 6, sobre esta clasificación existe claridad en un 69% de los estudiantes, por su parte un 27% reconoce la presencia de planetas y asteroides y un 4% entiende la posibilidad de una clasificación por predominancia de la materia en estado líquido.

Figura 6. Porcentajes de respuestas de los estudiantes a la Pregunta 4 sobre clasificación de los planetas

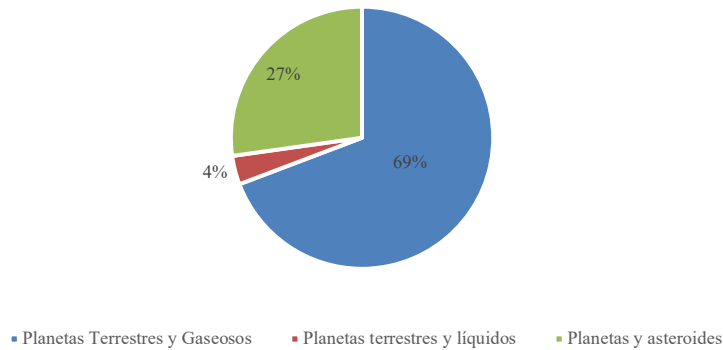

■ Planetas Terrestres y Gaseosos   ■ Planetas terrestres y líquidos   ■ Planetas y asteroides

Fuente: Elaboración propia.

Esta clasificación, así como las principales características del sistema solar, son precisas abordarlas en los grados 4° y 5° de educación primaria, además de ser denominados como el principal tema de diseño de material didáctico en Colombia (Valderrama & Navarrete Flórez, 2020).

Estrellas

El Sol es la única estrella del sistema solar, una estrella de tipo G de la secuencia principal y la fuente primaria de radiación electromagnética para la Tierra, con una energía aproximada de 1361 W/m². Para estudiar la composición química de las estrellas, en Astronomía se suele utilizar la espectroscopía, una técnica que analiza los espectros de emisión y absorción de luz (Clocchiatti & Catelan, 2017). Además, el Sol contiene una serie de regiones esféricas de acuerdo con el modelo de estructura solar vigente: el núcleo, la zona radiante, la zona convectiva, la fotósfera, la cromósfera y la corona (Cubas Armas, 2019). Estas afirmaciones son fundamentales para las siguientes preguntas.

*Pregunta 5*

La siguiente imagen (Figura 7) presenta la relación longitud de onda con región del espectro electromagnético. Empezando por la región con mayor longitud de onda; la opción que integra un mayor número de regiones espectrales es:

a) Radio, Infrarrojo, Luz visible, ultravioletas, rayos X, rayos gamma.
b) Colores Violeta, azul, verde, amarillo, naranja, rojo.
c) Rayos gamma, beta, x, ultravioletas, infrarrojos, ondas de radio.
d) Infrarrojo, rojo, naranja, amarillo, verde, azul, violeta y ultravioleta

Figura 7. Relación de la longitud de onda con el espectro electromagnético

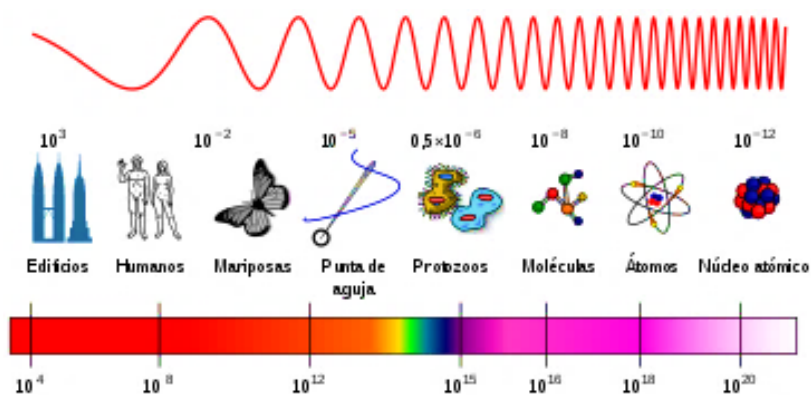

Fuente: File: EM Spectrum Properties es.svg - Wikimedia Commons.

En la Figura 8 se muestra que el 49% de los estudiantes identifican correctamente las regiones del espectro electromagnético según la longitud de onda; el 17% no reconoce que la radiación beta no forma parte de este espectro, y un 24% presenta dificultades para establecer dicha relación. Aunque esta pregunta se basa en temas tratados en los grados 8° y 9° de secundaria (Ministerio de Educación Nacional, 2011), se hace necesario abordar y fortalecer conceptos sobre el espectro electromagnético en los grados 6° y 7° para comprender los procesos de evolución estelar, como sugieren estudios realizados en otras regiones del país.

Figura 8. Porcentaje de respuestas frente a la pregunta 5 referente a las regiones del espectro electromagnético

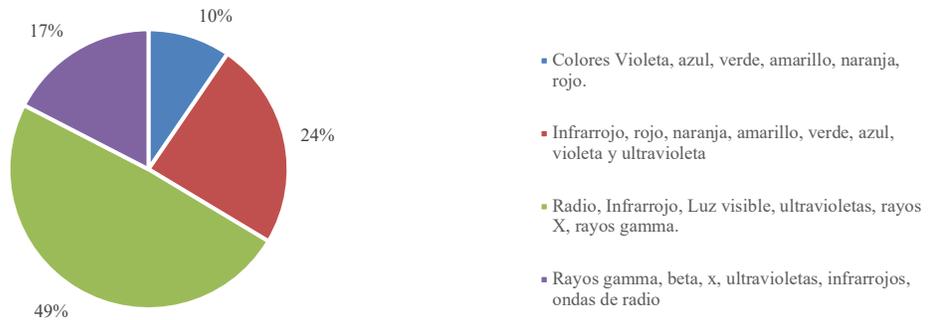

Fuente: Elaboración propia.

*Pregunta 6*

En una investigación científica se le solicita identificar los componentes químicos de la estrella Alfa-Centauro a partir de las mediciones de longitudes de onda del espectro de absorción y emisión, elija la técnica que usaría para realizar la investigación.

a) Espectrofotometría; técnica analítica utilizada para medir cuánta luz absorbe y emite una sustancia química.

b) Espectroscopia: técnica que utiliza los espectros de absorción y emisión para observar la interacción entre la luz y la materia

c) Sondas espaciales: consiste en enviar robots automatizados para colectar muestras, analizarlas y enviarlas de vuelta a la Tierra.

261

Figura 9. Porcentaje de respuestas sobre las técnicas de identificación de componentes químicos de una estrella

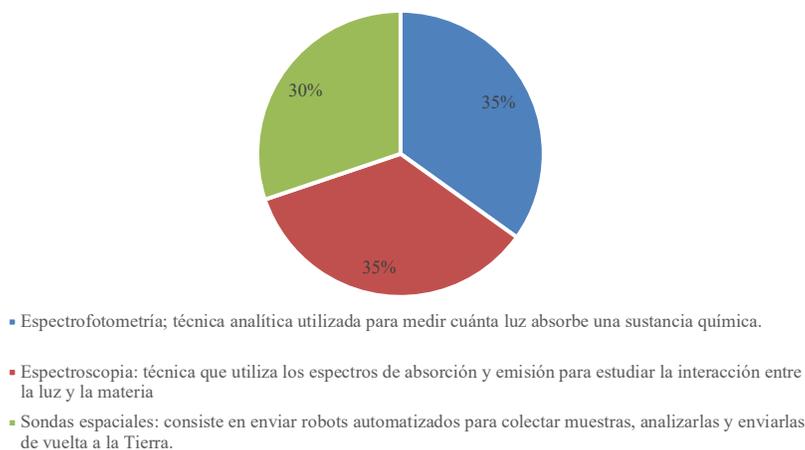

- Espectrofotometría; técnica analítica utilizada para medir cuánta luz absorbe una sustancia química.
- Espectroscopia: técnica que utiliza los espectros de absorción y emisión para estudiar la interacción entre la luz y la materia
- Sondas espaciales: consiste en enviar robots automatizados para colectar muestras, analizarlas y enviarlas de vuelta a la Tierra.

Fuente: Elaboración propia

La distribución porcentual de la Figura 9 presenta una proximidad entre las diferentes acciones investigadas. Esto puede sugerir que, si bien los estudiantes han visto algunos de los temas astronómicos revisados en esta investigación, a menudo se han obviado los métodos de estudio de la Astronomía. Esta situación podría ser problemática, ya que se estarían repitiendo conceptos astronómicos como verdades absolutas sin un entendimiento profundo de los métodos científicos involucrados.

Un 35% de los estudiantes asocia la espectrofotometría con el estudio de los componentes químicos de los astros de manera lógica, mientras que otro 35% cree que esto se realiza solo mediante observación, sin mediciones. Además, el 30% considera que es fácil enviar sondas espaciales a las estrellas. En revisiones anteriores, se ha señalado la escasez de investigaciones sobre procesos de evolución y estudio estelar.

Es necesario desarrollar estrategias educativas integradoras e interdisciplinarias que incluyan métodos científicos, análisis de compuestos químicos y fenómenos ondulatorios (Anzola-Triviño, 2021; Valderrama et al., 2021a).

*Pregunta 7*

De las siguientes opciones cuáles son regiones del Sol:

a) Núcleo, Manto superior, manto inferior, litosfera, hidrosfera

b) <u>Corona, cromosfera, fotosfera, núcleo</u>

c) Troposfera, estratosfera, mesosfera, exosfera

d) Núcleo, cromosfera, fotósfera y atmósfera

En la Figura 10, se observa que el 41% de los estudiantes identifica las capas del Sol y el 59% restante no tiene claridad frente al concepto, ya que un 29% relacionó la respuesta con las capas de la Tierra y no del Sol, así como un 5% indicó como capas solares las capas de la atmósfera terrestre, además un 25% no reconoce que tanto como la cromósfera y fotósfera son capas de la atmósfera solar, poniendo esta última como una capa adicional.

Figura 10. Porcentaje de respuestas frente a la pregunta 8 referente a las capas solares

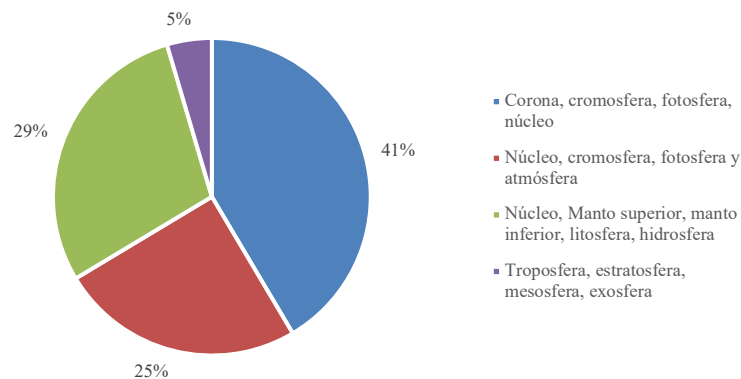

Fuente: Elaboración propia.

El abordaje conceptual de la estructura del Sol está presente en los Derechos Básicos de Aprendizaje para 6° y en los Estándares para 7° y 8°. Sin embargo, se reconoce la necesidad de comprender procesos más complejos de física y química relacionados con el Sol, especialmente la convección, la radiación y otros fenómenos, que se abordan en niveles educativos superiores. Por ende, sería conveniente reestructurar el currículo para organizar de manera coherente los temas astronómicos en relación con otros conceptos de Ciencias Naturales.

Por otra parte, en primaria, estos conceptos se abordan contextualizando los saberes y relacionándolos con los flujos de energía solar en los ecosistemas (Huérfano, 2013). Sin embargo, aportes como los realizados por Cutiva (2013) señalan que los docentes en formación presentan dificultades para aplicar conceptos físicos, matemáticos y herramientas computacionales en fenómenos astronómicos. Por lo tanto, surge la necesidad de adoptar una relación interdisciplinaria en la didáctica de las ciencias con la Astronomía.

*Pregunta 8*

Dándole continuidad a la categoría de Estrellas, se plantea la pregunta 8. Uno de los modelos actuales de clasificación de las estrellas es el modelo de Morgan-Keenan (Morgan & Kennan, 1973) el cual asigna clases a partir de la temperatura (color), tal como se muestra en la Figura 11. Si el Sol posee una temperatura de 5.500 K, se podría decir que pertenece a la categoría:

a) A y es una estrella Blanca
b) M y es roja
c) <u>G y es amarilla</u>
d) K y es naranja

Figura 11. Clasificación de las estrellas

| Clase | Temperatura[1] (Kelvin) | Color convencional |
|---|---|---|
| O | ≥ 33 000 K | azul |
| B | 10 000–33 000 K | azul a blanco azulado |
| A | 7500–10 000 K | blanco |
| F | 6000–7500 K | blanco amarillento |
| G | 5200–6000 K | amarillo |
| K | 3700–5200 K | naranja |
| M | ≤ 3700 K | rojo |

Fuente: Clasificación estelar, Wikipedia, la enciclopedia libre.

En la Figura 12, se puede observar que el 72% de los estudiantes enfrentan dificultades para relacionar la temperatura del Sol con la clasificación propuesta por el modelo Morgan-Keenan, mientras que el 28% logra establecer esta relación. Esto destaca la necesidad de una conceptualización previa que permita a los estudiantes comprender la tabla que están interpretando.

Polanco (2017) establece que estas dificultades se observan en diferentes niveles de formación como resultado del limitado acercamiento de los docentes a conceptos astronómicos, especialmente en el área de la astronomía estelar. Esto provoca la falta de clases sobre el tema y la escasez de contenidos astronómicos en los libros de consulta. Sin embargo, en otras regiones del país se están realizando esfuerzos para fortalecer estos conceptos a través de enfoques de formación docente, por lo que es necesario que en la región donde se llevó a cabo este estudio se exploren dichas posibilidades (Polanco, 2017; Vallejo V, 2022).

Figura 12. Porcentaje de respuestas frente a la pregunta 9 referente a la clasificación de las estrellas

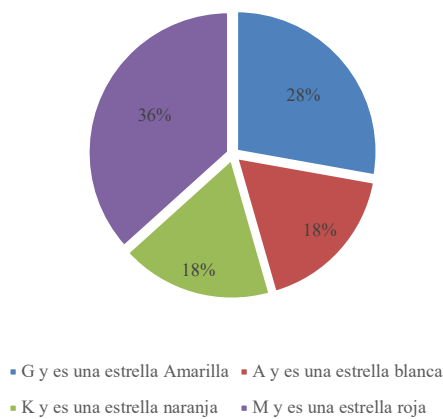

Fuente: Elaboración propia.

*Pregunta 9*

Los Estándares Básicos de Competencias para los grados sexto y séptimo plantean como indicadores de logro "describir el proceso de formación y extinción de las estrellas". Paralelo a esto, en Química se debe "describir los modelos que explican la estructura de la materia" y "explicar el desarrollo de modelos de organización de los elementos químicos" (Ministerio de Educación Nacional, 2011). Para alcanzar dichos objetivos, es fundamental entender la nucleosíntesis química, un proceso mediante el cual las partículas subatómicas se combinan para formar diferentes elementos atómicos. Según Escalante y Gasque (2012), existen cuatro tipos de nucleosíntesis química: Big Bang, estelar, supernovas y medio interestelar.
A partir de estos conceptos, se planteó la siguiente pregunta:

El ciclo de vida de las estrellas es importante en la comprensión del concepto de Nucleosíntesis, que es el proceso por medio del cual se forman los elementos químicos, los tipos de Nucleosíntesis existentes son:
a) Primordial, estelar, planetaria, satelital.
b) Nucleosíntesis del Big Bang, Nucleosíntesis planetaria, Nucleosíntesis de las supernovas y Nucleosíntesis del laboratorio.

c) <u>Nucleosíntesis del Big Bang, Nucleosíntesis estelar, Nucleosíntesis de las supernovas y Nucleosíntesis del medio interestelar.</u>

d) Nucleosíntesis del Big Bang, Nucleosíntesis estelar, Nucleosíntesis de las supernovas y Nucleosíntesis planetaria.

En la Figura 13, se observa que el 36% de los estudiantes lograron identificar los cuatro tipos de nucleosíntesis, mientras que el 63% presenta dificultades con el concepto. Esta dificultad puede deberse a la complejidad de estos procesos, los cuales pueden no ser pertinentes para los grados en los que se abordan estos temas. Además, es posible que estos conceptos se aborden de manera meramente histórica, sin un soporte adecuado en conocimientos fisicoquímicos para los estudiantes.

Por lo tanto, se debería introducir como mínimo el concepto de fusión nuclear para evitar concepciones erróneas sobre la producción de energía en las estrellas, la combustión del Sol y la formación de los elementos durante el Big Bang (Polanco-Erazo, 2017; Moya, 2019).

Figura 13. Porcentaje de respuestas frente a la pregunta 9 referente a los tipos de nucleosíntesis

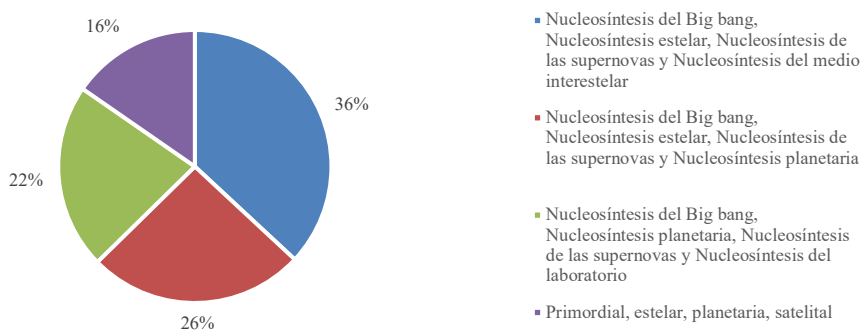

Fuente: Elaboración propia.

Galaxias y cosmología

La cosmología aborda el estudio de la composición, evolución y propiedades del universo con el objetivo de comprender su origen, desarrollo y posible futuro. Dentro de este campo, se

267

destacan conceptos importantes, como la posición del Sistema Solar a escala cósmica. A pesar de que la ciencia moderna enfrenta dificultades para definir un modelo específico del universo, es posible referirse a la posición relativa de este sistema dentro de la galaxia y a la posición de la galaxia con respecto a grupos de galaxias. Este planteamiento fundamenta la siguiente pregunta:

*Pregunta 10*

Con respecto a la Vía Láctea nuestro Sistema Solar se encuentra:

a) En el centro de la Vía Láctea
b) En la zona central de los brazos espirales
c) Fuera de la Vía Láctea
d) <u>En la zona externa de uno de los brazos espirales</u>

En la Figura 14, se observa que solo el 29% de los estudiantes ubica correctamente al sistema solar en el centro de uno de los brazos espirales de la Vía Láctea, el 32% lo ubica en el extremo, el otro 29% lo coloca en el centro de la Vía Láctea y un 10% lo sitúa por fuera de ella. Esto indica que el 70% de los estudiantes no reconoce con precisión la ubicación del sistema solar en la Vía Láctea.

Figura 14. Porcentaje de respuestas frente a la pregunta 10 referente a la ubicación del Sistema Solar en la Vía Láctea

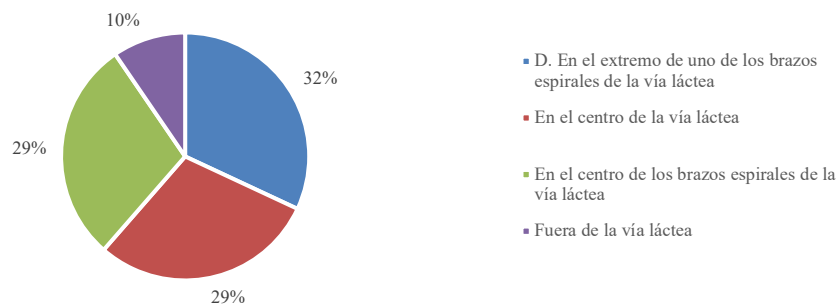

Fuente: Elaboración propia.

268

Los hallazgos revelan la carencia de ciertos contenidos astronómicos en los planes de estudio colombianos, especialmente conceptos relacionados con galaxias, conglomerados y cúmulos. Esto es significativo, ya que en la región es común observar una parte del disco galáctico, lo que podría confundir a los estudiantes sobre la ubicación del sistema solar. La falta de comprensión sobre la formación planetaria y las dinámicas galácticas podría mantener la creencia errónea de que estamos en el centro del universo o incluso en el centro de la galaxia.

A pesar de estas limitaciones, algunas áreas del país están implementando estrategias que integran estos conceptos, a menudo utilizando Tecnologías de la Información y la Comunicación (TIC) (González-Murillo, 2016). Inclusive, se han realizado iniciativas que permiten a los estudiantes interactuar con bases de datos astronómicos. Los resultados positivos de estas propuestas señalan la necesidad de incluir estos contenidos de manera explícita en el currículo y ampliar estas prácticas a nivel nacional (Cruz, 2020).

*Pregunta 11*

En términos cosmológicos, la materia en el universo está regulada por cuatro interacciones fundamentales, que incluyen la gravedad, el electromagnetismo y las interacciones nucleares. Pese a que el currículo colombiano no aborda la física y química modernas o contemporáneas, se buscó explorar estas interacciones fundamentales con la siguiente pregunta.

La materia en el universo interactúa por medio de las siguientes cuatro fuerzas fundamentales:
a) La gravedad, el electromagnetismo, solar y nuclear
b) La gravedad, el electromagnetismo, química y física
c) La gravedad, el electromagnetismo, motora y de fricción
d) La gravedad, el electromagnetismo, fuerza nuclear débil y nuclear fuerte

Del análisis de la Figura 15, se observa que el 33% de los estudiantes identifica correctamente a la gravedad, el electromagnetismo, la fuerza débil y la fuerza fuerte como las cuatro fuerzas fundamentales. Por otro lado, un 28% menciona la gravedad, el electromagnetismo, la fuerza solar y la fuerza nuclear como las cuatro fuerzas. Un 20% indica que son la gravedad, el electromagnetismo, la química y la física, mientras que el 19% restante señala que las cuatro fuerzas son la gravedad, el electromagnetismo, la fuerza motriz y la de fricción. Esto revela que el 67% de los estudiantes no tienen claridad frente al concepto de las interacciones de la naturaleza.

Figura 15. Porcentaje de respuestas frente a la pregunta 11 referente a las cuatro fuerzas de la naturaleza

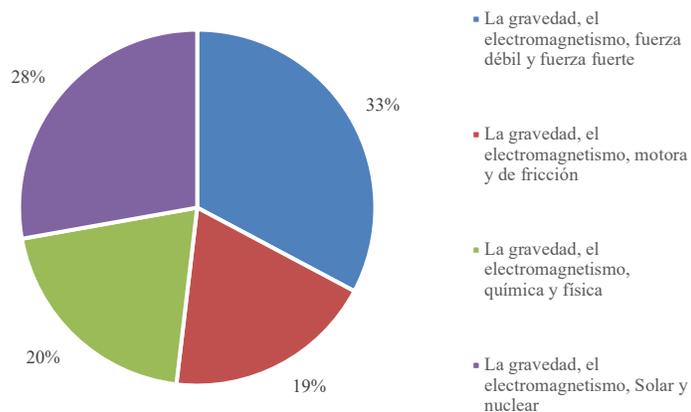

Fuente: Elaboración propia.

La complejidad del concepto justifica en parte su desconocimiento. Sin embargo, es crucial que la educación secundaria aborde estos contenidos desde la física, integrando conceptos que respaldan el conocimiento físico y astronómico de manera integral, donde se contextualizan estos conceptos a la realidad de los estudiantes, presentando las ciencias naturales, incluida la astronomía, como un todo, promoviendo una visión de complementariedad entre estas, lo que podría fomentar una actitud de aplicabilidad y reducir la percepción de abstracción y dificultad en estas ciencias.

**Conclusiones**

Se puede concluir que los estudiantes muestran ciertos conocimientos previos en Astronomía, especialmente sobre las fases de la Luna y aspectos del Sistema Solar, en línea con el currículo

270

establecido. Sin embargo, la comprensión de los movimientos de la Tierra necesita más profundización, especialmente en relación con los patrones climáticos. Por lo tanto, es esencial desarrollar estrategias didácticas que refuercen conceptos científicos, vinculándolos con otras disciplinas como Física, Química y Biología.

Asimismo, se hace necesario revisar y actualizar la distribución de los conceptos astronómicos en los lineamientos curriculares nacionales para asegurar que estén alineados con la progresión de la comprensión de otros conceptos físicos y químicos que sustentan la Astronomía. Esto podría fomentar una comprensión más profunda del conocimiento científico y promover la participación en carreras científicas, el desarrollo tecnológico y las dinámicas sociales.

**Referencias**